\documentclass[utf8]{FrontiersinHarvard} 

\usepackage{url,hyperref,microtype,subcaption, here}
\usepackage[onehalfspacing]{setspace}

\def\keyFont{\fontsize{8}{11}\helveticabold }
\def\firstAuthorLast{Ishikawa {et~al.}} 
\def\Authors{Yuto Ishikawa\,$^{1,*}$, Takuma Yoshihara\,$^{2,*}$, Keita Okamura\,$^{3,*}$ and Masayuki Ohzeki\,$^{4,5,6}$}
\def\Address{$^{1}$Department of Computer Science, Nagoya University, Nagoya, Aichi, Japan \\
$^{2}$Department of Engineering, Tokyo Denki University, Adachi, Tokyo, Japan \\
$^{3}$Department of Physics and Astronomy, Tokyo University of Science, Noda, Chiba, Japan\\
$^{4}$Graduate School of Information Sciences, 
  Tohoku University
  Sendai, Miyagi, Japan \\
$^{5}$Department of Physics, Tokyo Institute of Technology, Meguro, Tokyo, Japan \\
$^{6}$Sigma-i Co., Ltd.,
  Shinagawa, Tokyo, Japan }

\def\corrAuthor{Corresponding Author}

\def\corrEmail{
ishikawa.yuto.f6@s.mail.nagoya-u.ac.jp\\
21ef106@ms.dendai.ac.jp\\
6223024@ed.tus.ac.jp}

\usepackage{url,hyperref,lineno,microtype,subcaption}

\begin{document}
\onecolumn
\firstpage{1}

\title[Quantum Annealing Based Maze Generation]{Individual subject evaluated difficulty of adjustable mazes generated using quantum annealing} 

\author[\firstAuthorLast ]{\Authors} 
\address{} 
\correspondance{} 

\extraAuth{}

\maketitle

\begin{abstract}
\section{}
    In this paper, the maze generation using quantum annealing is proposed. We reformulate a standard algorithm to generate a maze into a specific form of a quadratic unconstrained binary optimization problem suitable for the input of the quantum annealer. To generate more difficult mazes, we introduce an additional cost function $Q_{update}$ to increase the difficulty. The difficulty of the mazes was evaluated by the time to solve the maze of 12 human subjects.
    To check the efficiency of our scheme to create the maze, we investigated the time-to-solution of a quantum processing unit, classical computer, and hybrid solver.

\tiny
 \keyFont{ \section{Keywords:} quantum annealing, combinatorial optimization, maze generation, bar-tipping algorithm, time-to-solution}
\end{abstract}

\section{Introduction}
A combinatorial optimization problem is minimizing or maximizing their cost or objective function among many variables that take discrete values.
In general, it takes time to solve the combinatorial optimization problem.
To deal with many combinatorial optimization problems, we utilize generic solvers to solve them efficiently.
Quantum annealing (QA) is one of the generic solvers for solving combinatorial optimization problems \cite{Kadowaki1998} using the quantum tunneling effect.
Quantum annealing is a computational technique to search for good solutions to combinatorial optimization problems by expressing the objective function and constraint time requirements of the combinatorial optimization problem by quantum annealing in terms of the energy function of the Ising model or its equivalent QUBO (Quadratic Unconstrained Binary Optimization), and manipulating the Ising model and QUBO to search for low energy states \cite{Tanaka2022}.
Various applications of QA are proposed in traffic flow optimization \cite{neukart2017traffic,hussain2020optimal,inoue2021traffic}, 
 finance \cite{rosenberg2016solving, orus2019forecasting, venturelli2019reverse}, logistics \cite{feld2019hybrid,ding2021implementation}, manufacturing \cite{venturelli2016quantum, Yonaga2022, Haba2022}, preprocessing in material experiments \cite{Tanaka2023}, marketing \cite{nishimura2019item}, steel manufacturing \cite{Yonaga2022}, and decoding problems \cite{IdeMaximumLikelihoodChannel2020, Arai2021code}.
The model-based Bayesian optimization is also proposed in the literature \cite{Koshikawa2021}
A comparative study of quantum annealer was performed for benchmark tests to solve optimization problems \cite{Oshiyama2022}. 
The quantum effect on the case with multiple optimal solutions has also been discussed \cite{Yamamoto2020, Maruyama2021}. 
As the environmental effect cannot be avoided, the quantum annealer is sometimes regarded as a simulator for quantum many-body dynamics \cite{Bando2020, Bando2021, King2022}. 
Furthermore, applications of quantum annealing as an optimization algorithm in machine learning have also been reported \cite{neven2012qboost,khoshaman2018quantum,o2018nonnegative, Amin2018,Kumar2018,Arai2021,Sato2021,Urushibata2022,hasegawa2023,Goto2023}.
In this sense, developing the power of quantum annealing by considering hybrid use with various techniques is important, as in several previous studies \cite{Hirama2023, Takabayashi2023}.

In this study, we propose the generation of the maze by quantum annealing.
In the application of quantum annealing to mazes, algorithms for finding the shortest path through a maze have been studied \cite{Pakin2017}.
Automatic map generation is an indispensable technique for game production, including roguelike games. Maze generation has been used to construct random dungeons in roguelike games, by assembling mazes \cite{Cheong-mok2015}.
Therefore, considering maze generation as one of the rudiments of this technology, we studied maze generation using a quantum annealing machine.
Several algorithms for the generation of the maze have been proposed.
In this study, we focused on maze-generating algorithms.
One can take the bar-tipping algorithm \cite{Algoful_MazeBar}, the wall-extending algorithm \cite{Algoful_Extend}, and the hunt-and-kill algorithm \cite{Algoful_MazeDig}.
    
\begin{figure}[h!]
\begin{center}
\includegraphics[width=6.5cm]{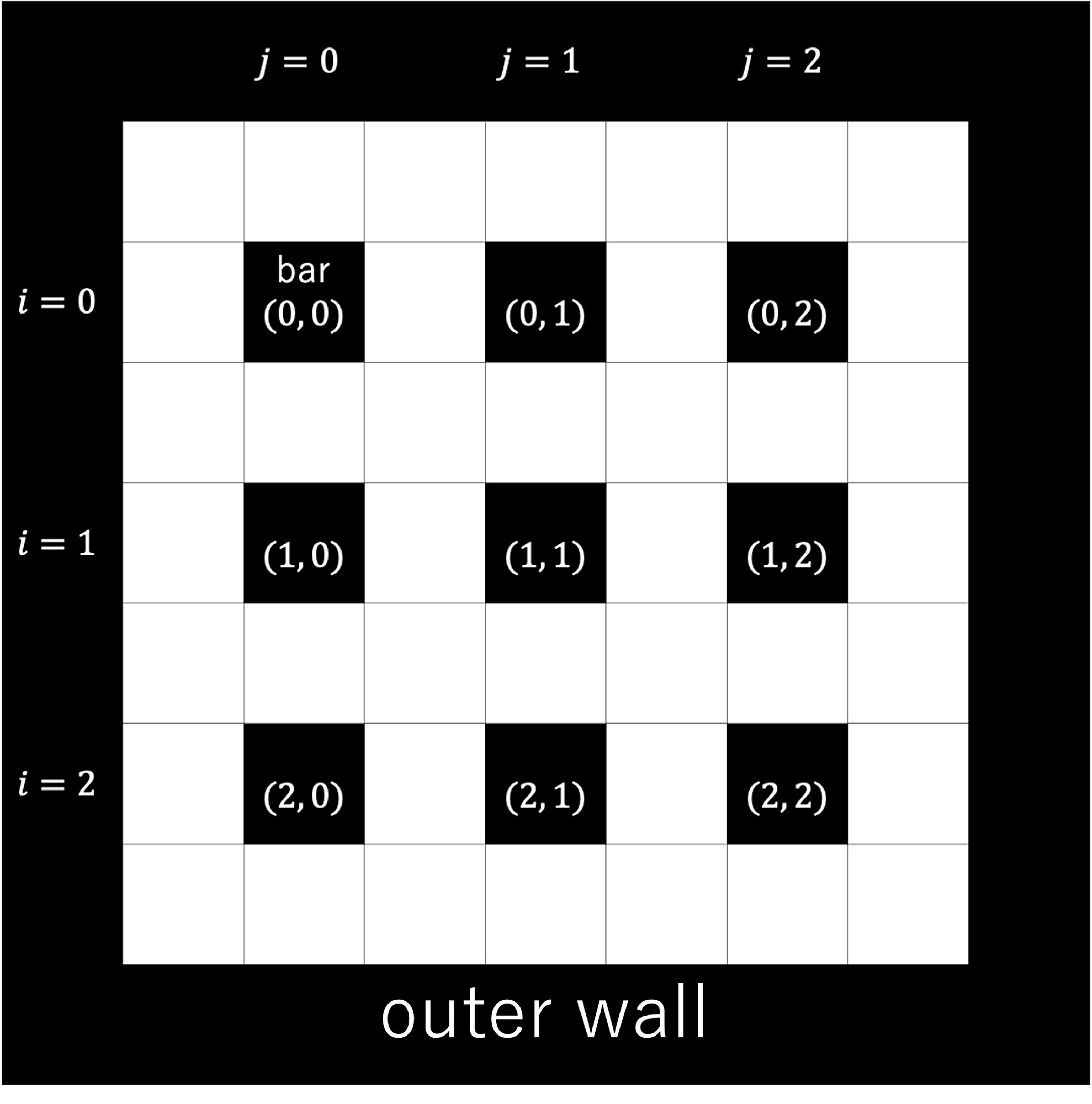}
\end{center}
\caption{ Positions where outer wall, bars, and coordinate $(i, j)$ are in $3\times3$ maze. 
}\label{Bar-tipping_algorithm_step0}
\end{figure}

\begin{figure}[h!]
\begin{center}
\includegraphics[width=15cm]{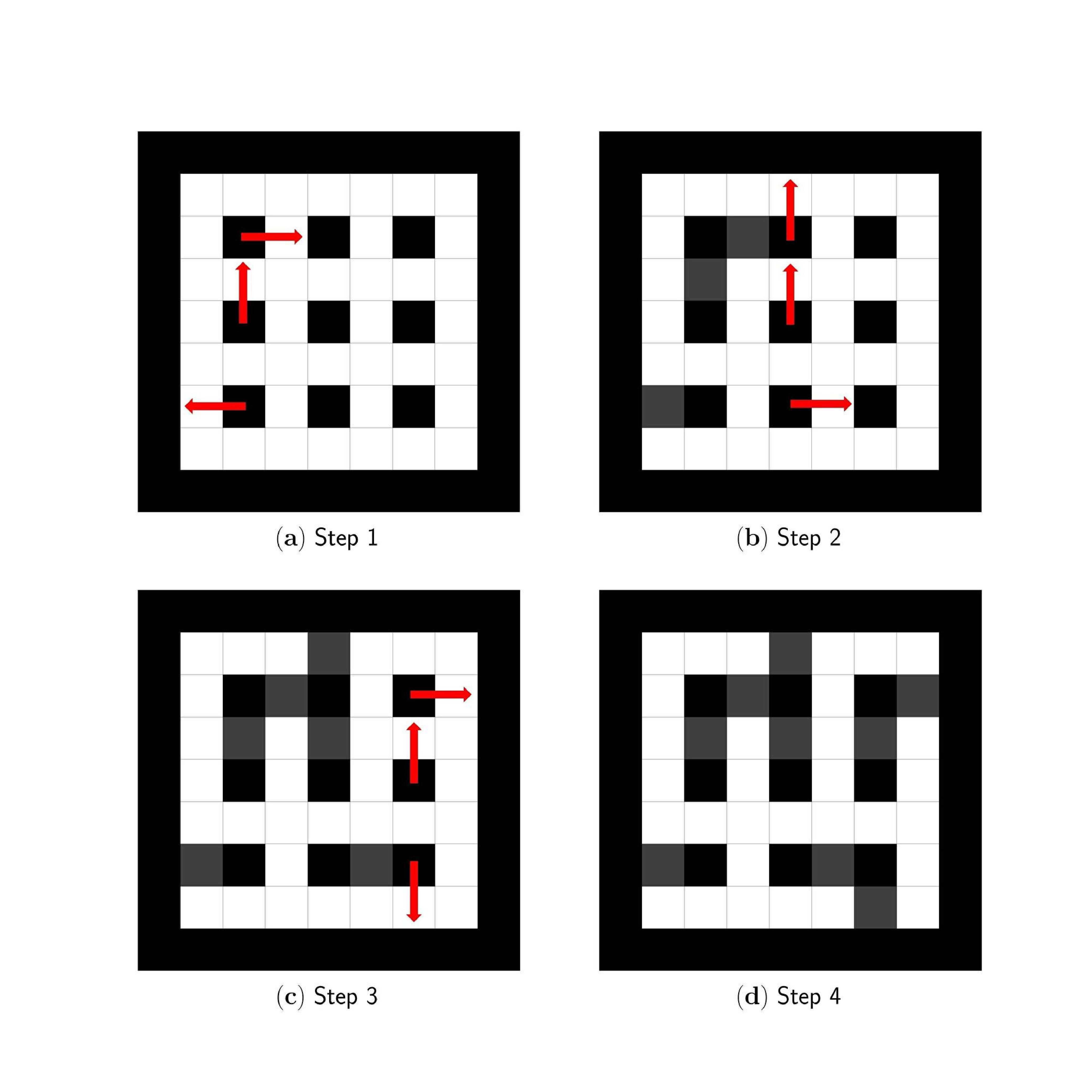}
\end{center}
\caption{ Step of bar-tipping algorithm. \textbf{(a)} step1: bars in first column are extended. \textbf{(b)} step2: bars in second column are extended. \textbf{(c)} step3: bars in subsequent column are extended. \textbf{(d)} step4: A complete maze through these steps.
}\label{Bar-tipping_algorithm}
\end{figure}

    The bar-tipping algorithm is an algorithm that generates a maze by extending evenly spaced bars one by one. 
    For the sake of explanation, we will explain the terminology here. A path represents an empty traversable part of the maze and a bar a filled non traversable part.
    Figure \ref{Bar-tipping_algorithm_step0} shows where the outer wall, bars, and coordinate $(i, j)$ are in a $3\times3$ maze. The maze is surrounded by an outer wall as in Figure \ref{Bar-tipping_algorithm_step0}.
    It requires the following three constraints. 
    First, each bar can be extended by one cell only in one direction. Second, the first column can be extended in four directions: up, down, left, and right, while the second and subsequent columns can be extended only in three directions: up, down, and right. Third, adjacent bars cannot overlap each other. We explain the detailed process of the bar-tipping algorithm using the $3\times3$ size maze. 
    In this study, a maze generated by extending the \(N \times N\) bars is called \(N \times N\) size maze.  
    First, standing bars are placed in every two cells in a field surrounded by an outer wall, as in Figure \ref{Bar-tipping_algorithm_step0}. 
    Second, Figure \ref{Bar-tipping_algorithm} shows each step of bar-tipping algorithm. 
    Figure \ref{Bar-tipping_algorithm} (a) shows the first column of bars extended.
    The bars in the first column are randomly extended in only one direction with no overlaps, as in Figure \ref{Bar-tipping_algorithm} (a). 
    The bars can be extended in four directions (up, down, right, left) at this time. 
    Figure \ref{Bar-tipping_algorithm} (b) shows the second column of bars being extended. Third, the bars in the second column are randomly extended in one direction without overlap as in Figure \ref{Bar-tipping_algorithm} (b). 
    The bars can be extended in three directions (up, down, right) at this time. Figure \ref{Bar-tipping_algorithm} (c) shows the state in which the bars after the second column are extended.
    Fourth, the bars in subsequent columns are randomly extended in one direction, likewise the bars in the second column, as in Figure \ref{Bar-tipping_algorithm} (c). 
    Figure \ref{Bar-tipping_algorithm} (d) shows the complete maze in its finished state. Following the process, we can generate a maze as in Figure \ref{Bar-tipping_algorithm} (d).

\begin{figure}[h!]
\begin{center}
\includegraphics[width=15cm]{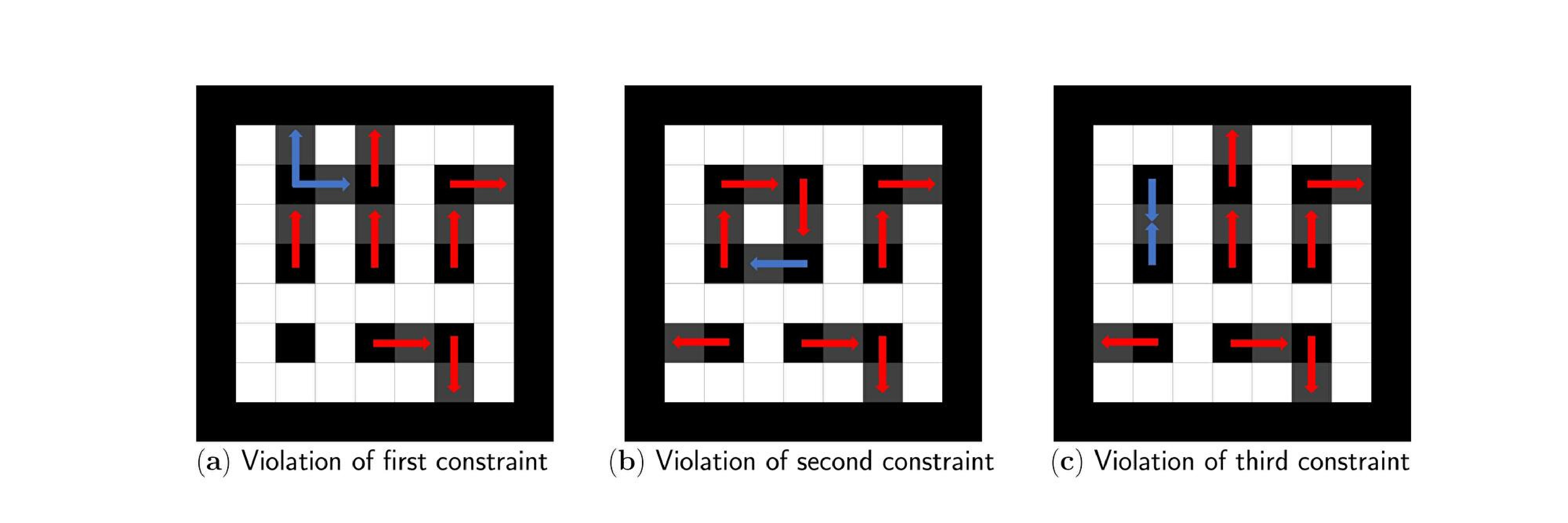}
\end{center}
\caption{ Mazes violated the constraints. \textbf{(a)} A maze violate the first constraint. \textbf{(b)} A maze violate the second constraint. \textbf{(c)} A maze violated the third constraint.
}\label{Bar-tipping_algorithm_constraints}
\end{figure}

If multiple maze solutions are possible, the maze solution is not unique, simplifying the time and difficulty of reaching the maze goal. 
These constraints must be followed for the reasons described below.
The first constraint prevents a maze from generating a maze with multiples maze solutions and closed circuits.
Figure \ref{Bar-tipping_algorithm_constraints} (a) shows a maze state that violates the first constraint.
The step violating the first constraint because one bar in the upper right corner is extended in two directions as Figure \ref{Bar-tipping_algorithm_constraints} (a) .

The second constraint prevents generating a maze from a maze with closed circuits and multiple maze solutions. 
Figure \ref{Bar-tipping_algorithm_constraints} (b) shows a state that violates the second constraint. 
The second constraint is violated, it has a closed circuit and multiple maze solutions, as  Figure \ref{Bar-tipping_algorithm_constraints} (b).
The third constraint prevents maze generation from a maze with multiple maze solutions. Figure \ref{Bar-tipping_algorithm_constraints} (c) shows a state that violates the third constraint. 
The bars overlap in the upper right corner, making it the third constraint as Figure \ref{Bar-tipping_algorithm_constraints} (c).

\begin{figure}[h!]
\begin{center}
\includegraphics[width=6.5cm]{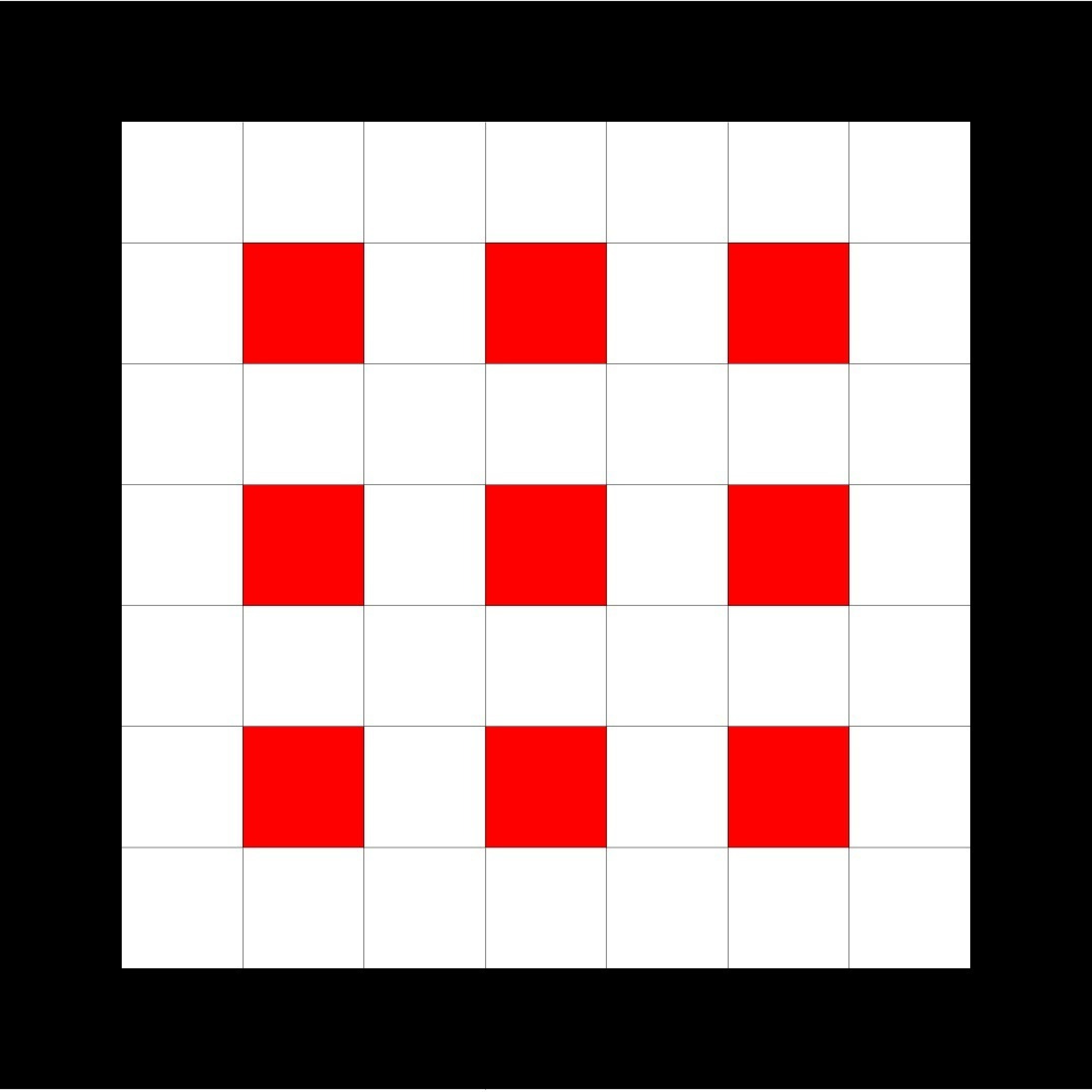}
\end{center}
\caption{ Red cells represent options of starting coordinates for the wall-extending algorithm.
}\label{Wall-extending_starting_coordinate}
\end{figure}

\begin{figure}[h!]
\begin{center}
\includegraphics[width=15cm]{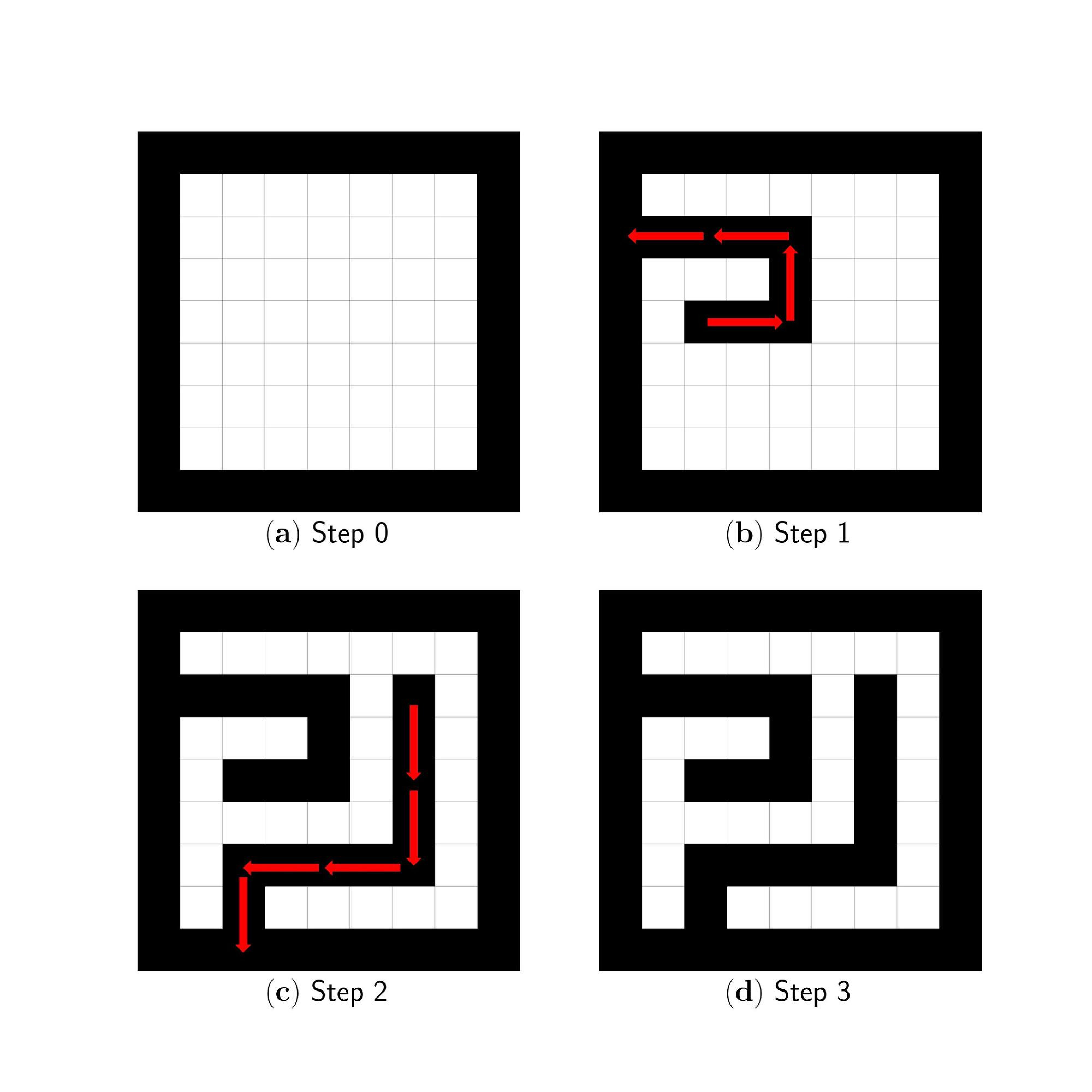}
\end{center}
\caption{ \textbf{(a)} Initial state for wall-extending algorithm. \textbf{(b)} Step 1 for wall-extending algorithm. \textbf{(c)} Step 2 for wall-extending algorithm. \textbf{(d)} Maze generated using wall-extending algorithm.
}\label{Wall-extending_algorithm}
\end{figure}

   Next, we describe the wall-extending algorithm. It is an algorithm that generates a maze by extending walls.
    Figure \ref{Wall-extending_starting_coordinate} shows the extension starting coordinates of the wall-extending algorithm.
    Figure \ref{Wall-extending_algorithm} (a) shows the initial state of the wall expansion algorithm.
    First, as an initial condition, the outer perimeter of the maze is assumed to be the outer wall, and the rest of the maze is assumed to be the path as  Figure \ref{Wall-extending_algorithm} (a). 
    Coordinate system is different from the bar-tipping algorithm, all cells are labeled coordinates.
    As Figure \ref{Wall-extending_starting_coordinate} shows, the coordinates where both $x$ and $y$ are even and not walls are listed as starting coordinates for wall extending.
  The following process is repeated until all starting coordinates change to walls, as shown in Figure 5(c). 
Randomly choose the coordinates from the non-wall extension start coordinates.

    The next extending direction is randomly determined from which the adjacent cell is a path.
    Figure \ref{Wall-extending_algorithm} (b) shows how the path is extended.
  The extension will be repeated while two cells ahead of the extending direction to be extended is a path as  Figure \ref{Wall-extending_algorithm} (b).
      Figure \ref{Wall-extending_algorithm} (c) shows all starting coordinates changed to walls.
    These processes are repeated until all the starting coordinates change to walls as in Figure \ref{Wall-extending_algorithm} (c).
    Figure \ref{Wall-extending_algorithm} (d) shows a maze created by wall-extending.
    Following the process, we can generate a maze as in Figure \ref{Wall-extending_algorithm} (d).

\begin{figure}[h!]
\begin{center}
\includegraphics[width=6.5cm]{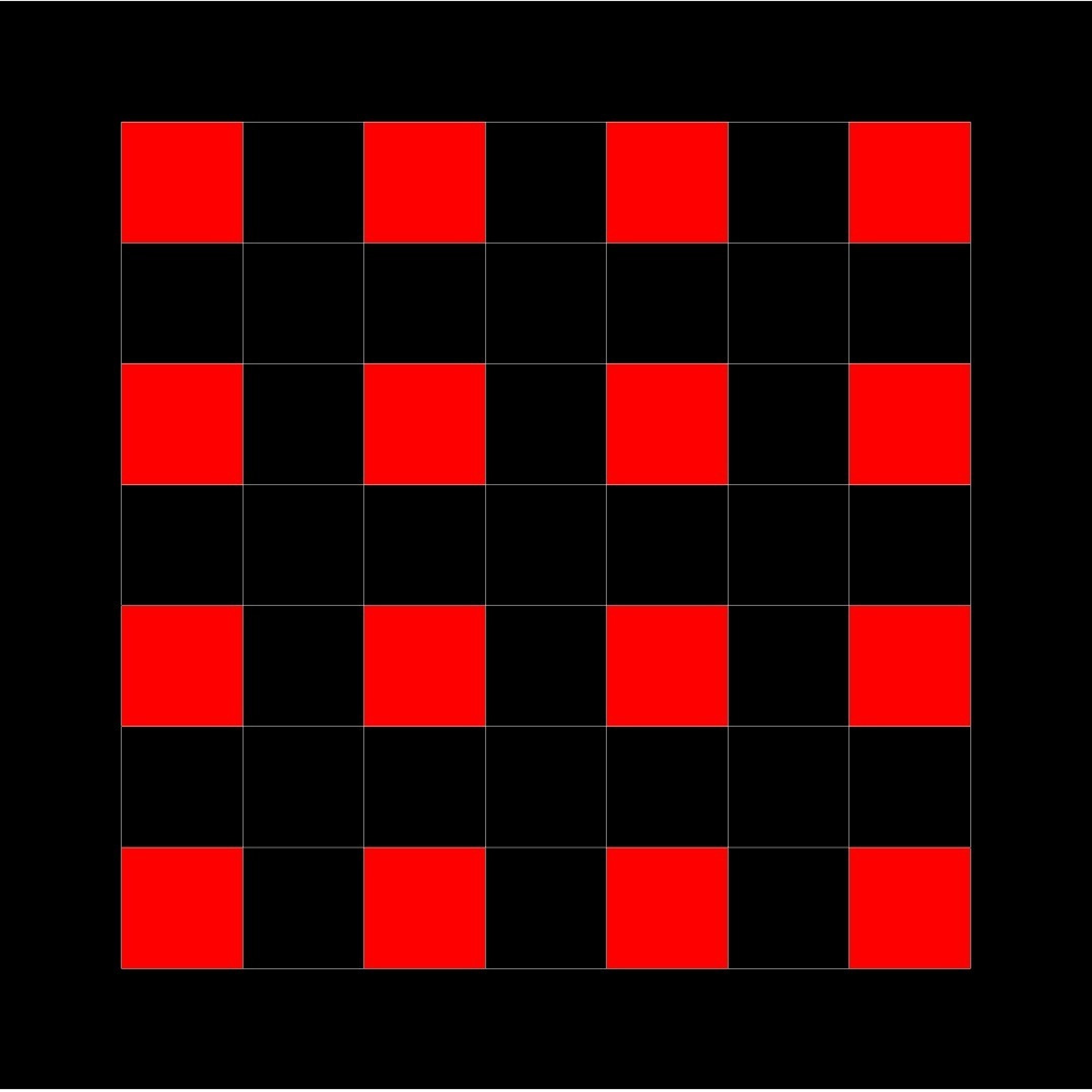}
\end{center}
\caption{ Red cells represent options of starting coordinates for the hunt-and-kill algorithm.
}\label{Hunt-and-kill_starting_coordinate}
\end{figure}

\begin{figure}[h!]
\begin{center}
\includegraphics[width=15cm]{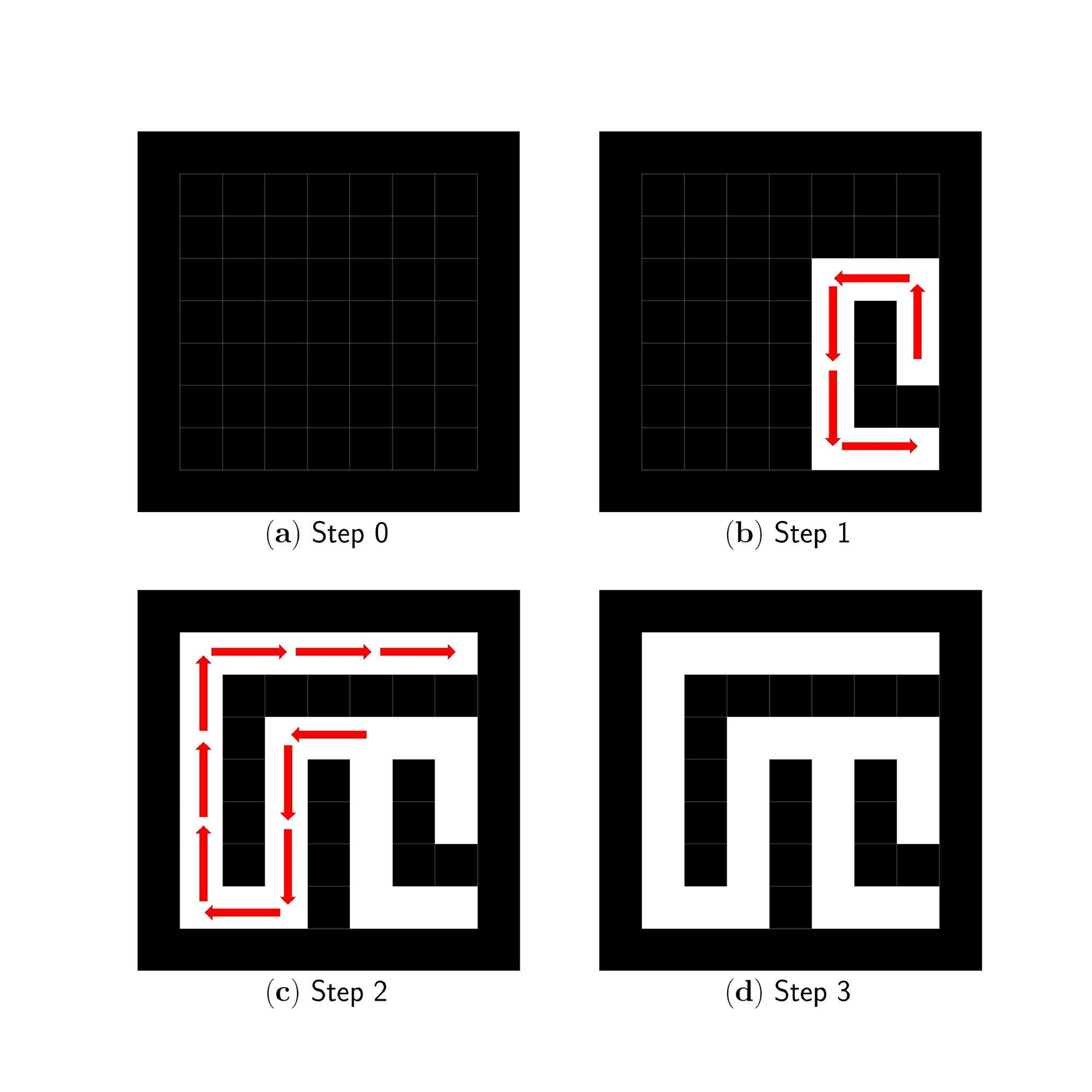}
\end{center}
\caption{ \textbf{(a)} Initial state for hunt-and-kill algorithm. \textbf{(b)} Step 1 for hunt-and-kill algorithm. \textbf{(c)} Step 2 for hunt-and-kill algorithm. \textbf{(d)} Maze generated using hunt-and-kill algorithm.
}\label{Hunt-and-kill_algorithm}
\end{figure}

     As a third, the hunt-and-kill algorithm is explained below. It is an algorithm that generates a maze by extending paths. 
     Figure \ref{Hunt-and-kill_starting_coordinate} shows the extension starting coordinates of the hunt-and-kill algorithm.
 Figure \ref{Hunt-and-kill_algorithm} (a) shows the initial state of the hunt-and-kill algorithm. The entire surface is initially walled off as Figure \ref{Hunt-and-kill_algorithm} (a). 
     Coordinates, where both $x$ and $y$ are odd, are listed as starting coordinates for path extension as in Figure \ref{Hunt-and-kill_starting_coordinate}. 
     As with the wall-extending algorithm, all cells are set to coordinates.
     Figure  \ref{Hunt-and-kill_algorithm} (b) shows the state in which the path is extended. A coordinate is chosen randomly from the starting coordinates, and the path is extended from there as in Figure \ref{Hunt-and-kill_algorithm} (b).
     Figure \ref{Hunt-and-kill_algorithm} (c) shows the coordinate selection and re-extension after the path can no longer be extended. If the path can no longer be extended, a coordinate is randomly selected from the starting coordinates, which are already paths, and extension starts again from it as in Figure \ref{Hunt-and-kill_algorithm} (c).
     This process is repeated until all the starting coordinates turn into paths to generate the maze.
   Figure \ref{Hunt-and-kill_algorithm} (d) shows the complete maze with the hunt-and-kill algorithm. Following the process, we can generate a maze as in Figure \ref{Hunt-and-kill_algorithm} (d).

    Of the three maze generation algorithms mentioned above, the bar-tipping algorithm is relevant to the combinatorial optimization problem. 
In addition, unlike other maze generation algorithms, the bar-tipping algorithm is easy to apply because it only requires the consideration of adjacent elements. 
Thus, we have chosen to deal with this algorithm.
Other maze generation algorithms could be generalized by reformulating them as combinatorial optimization problems.
The wall-extending and hunt-and-kill algorithms will be implemented in future work, considering the following factors.
The former algorithm introduces the rule that adjacent walls are extended and so are their walls.
The number of connected components will be computed for the latter, and the result will be included in the optimization.
    
Using the bar-tipping algorithm, we reformulated it to solve a combinatorial optimization problem that generates a maze with a longer solving time and optimized it using quantum annealing.
Quantum annealing (DW\_2000Q\_6 from D-Wave), classical computing (simulated annealing, simulated quantum annealing, and algorithmic solution of the bar-tipping algorithm), and hybrid computing were compared with each other according to the generation time of mazes, and their performance was evaluated. 
The solver used in this experiment is as follows: DW\_2000Q\_6 from D-Wave, simulated annealer called SASampler and simulated quantum annealer called SQASampler from OpenJij \cite{openJij}, D-Wave's quantum-classical hybrid solver called hybrid\_binary\_quadratic\_model\_version2 (BQM) and classical computer (MacBook Pro(14-inch, 2021), OS: macOS Monterey Version 12.5, Chip: Apple M1 Pro, Memory: 16GB) 
This comparison showed that quantum annealing was faster. This may be because the direction of the bars is determined at once using quantum annealing, which is several times faster than the classical algorithm.
We do not use an exact solver to solve the combinatorial optimization problem.
We expect some diversity in the optimal solution and not only focus on the optimal solution in maze generation.
Thus, we compare three solvers, which generate various optimal solutions.

    In addition, we generate mazes that reflect individual characteristics, whereas existing maze generation algorithms rely on randomness and fail to incorporate other factors.
    In this case, we incorporated the maze solution time as one of the other factors to solve the maze.
The maze solving time was defined as the time (in seconds) from the start of solving the maze to the end of solving the maze.

    The paper is organized as follows.
    In the next Section, we explain the methods of our experiments. In Sec. 3, we describe the results of our experiments. In Sec. 4, we summarize this paper.

\section{Methods}
\subsection{Cost function}
To generate the maze by quantum annealer, we need to set the cost function in the quantum annealer.
One of the important features of the generation of the maze is diversity.
In this sense, the optimal solution is not always unique.
Since it is sufficient to obtain a structure consistent with a maze, the cost function is mainly derived from the necessary constraints of a maze, as explained below.
Three constraints describe the basis of the algorithm of the bar-tipping algorithm.
The cost function will be converted to a QUBO matrix to use the quantum annealer.To convert the cost function to a QUBO, the cost function must be written in a quadratic form.
Using the penalty method, we can convert various constraints written in a linear form into a quadratic function. The penalty method is a method to rewrite the equality constant as a quadratic function. For example, the penalty method can rewrite an equation constant $x = 1$ to $(x-1)^2$. 
Thus, we construct the cost function for generating the maze using the bar-tipping algorithm below. 

The constraints of the bar-tipping algorithm correlate with each term in the cost function described below.
The first constraint of the bar-tipping algorithm is that the bars can be extended in only one direction. It prevents making closed circuits.
The second constraint of the bar-tipping algorithm is that the bars of the first column be extended randomly in four directions (up, right, down, and left), and the second and subsequent columns can be extended randomly in three directions (up, right, and down). 
 It also prevents the creation of closed circuits.
 The third constraint of the bar-tipping algorithm is that adjacent bars must not overlap.
Following the constraint in the bar-tipping algorithm, we can generate a maze with only one path from the start to the goal. 

The cost function consists of three terms to reproduce the bar-tipping algorithm according to the three constraints and to determine the start and goal.
\begin{equation}
\begin{split}
E(\{x_{i,j,d},X_{m,n}\})
= \sum_{i,i'}\sum_{j,j'}\sum_{d,d'}Q_{(i, j, d), (i', j', d')}x_{i, j, d}x_{i', j', d'} + \lambda_1\sum_{i}\sum_{j}\Biggl(\sum_{d} x_{i, j, d} - 1\Biggr)^2
\\+ \lambda_2\Biggl(\sum_m\sum_nX_{m,n}-2\Biggr)^2,
\label{cost_function}
\end{split}
\end{equation}
where \(x_{i, j, d}\) denotes whether the bar in \(i\)-th row, \(j\)-th column extended in direction \(d~( {\rm up} \colon 0, {\rm right}\colon 1, {\rm down}\colon 2, {\rm left}\colon 3)\). When the bar in coordinate $(i,j)$ is extended in direction, $x_{i,j,d}$ takes $1$, otherwise takes $0$.
Due to the second constraint of the bar-tipping algorithm, the bars after the second column cannot be extended on the left side; only the first column has \((d=3)\).
Furthermore, $Q_{(i,j,d)(i',j',d')}$ in Equation \ref{cost_function} depends on $i, j, d, i', j'$, and $d'$ and is expressed as follows
\begin{equation}
\ Q_{(i, j, d), (i', j', d')}=
\left\{
\begin{array}{ll}
1&(i=i'-1, j=j', d=2, d'=0) \\
1&(i=i'+1, j=j', d=0, d'=2) \\
0&{\rm otherwise}.
\end{array}
\right.
\end{equation}
 The coefficients of $\lambda_{1}$ and $\lambda_{2}$ are constants to adjust the effects of each penalty term. The first term prevents the bars from overlapping and extending each other face-to-face. It represents the third constraint of the bar-tipping algorithm. Here, due to the second constraint, bars in the second and subsequent columns cannot be extended to the left. Therefore, the adjacent bars in the same row cannot extend and overlap. This corresponds to the fact that $d$ cannot take $3$ when $j \geq 1$. Thus, there is no need to reflect, considering the left and right. In particular, the first term restricts the extending and overlapping between the up and down adjacent bars. For example, the situation in which one bar in $(i, j)$ extended down $(d = 2)$ and the lower bar in $(i + 1, j)$ extended up $(d = 0)$ is represented by $x_{i,j,0}x_{i+1,j,2} =1$, and $Q{(i,j,0),(i+1,j,2)}$ takes $1$. In the same way, thinking of the relation between the bar in $(i,j)$ and the upper bar in $(i-1,j)$, $Q_{(i-1,j,2),(i,j,0)}=1$. Thus, $Q_{(i-1,j,2),(i,j,0)}x_{i,j,0}x_{i+1,j,2}$ takes 1, and the value of the cost function taken will increase. By doing this, the third constraint is represented as a first term. The second term is a penalty term that limits the direction of extending to one per bar. It represents the first constraint of the bar-tipping algorithm.
This means that for a given coordinate $(i,j)$, the sum of $x_{i,j,d}$ $\bigl(d=0,1,2(,3)\bigr)$ must take the value $1$. Here, the bars in the second and subsequent columns cannot extend to the left by the second constraint. Thus, $d$ takes (0, 1, 2, 3) when $j=0$, and $d$ takes (0, 1, 2) when $j\geq1$.
The third term is the penalty term for selecting two coordinates of the start and the goal from the coordinates $(m,n)$. This means that a given coordinate $(m,n)$, the sum of $X_{m,n}$ takes $2$.
The start and the goal are commutative in the maze. 
They are randomly selected from the two coordinates determined by the third term.
$X_{m,n}$ denotes whether or not to set the start and goal at the $m$-th row and $n$-th column of options of start and goal coordinates. When the $(m,n)$ coordinate is chosen as the start and goal, $X_{m,n}$ takes $1$. Otherwise, it takes $0$.
There are no relations between $X_{m,n}$ and $x_{i,j,d}$ in Equation \ref{cost_function}. This means that the maze structure and the start and goal determination coordinates have no relations.
\begin{figure}[h!]
\begin{center}
\includegraphics[width=6.5cm]{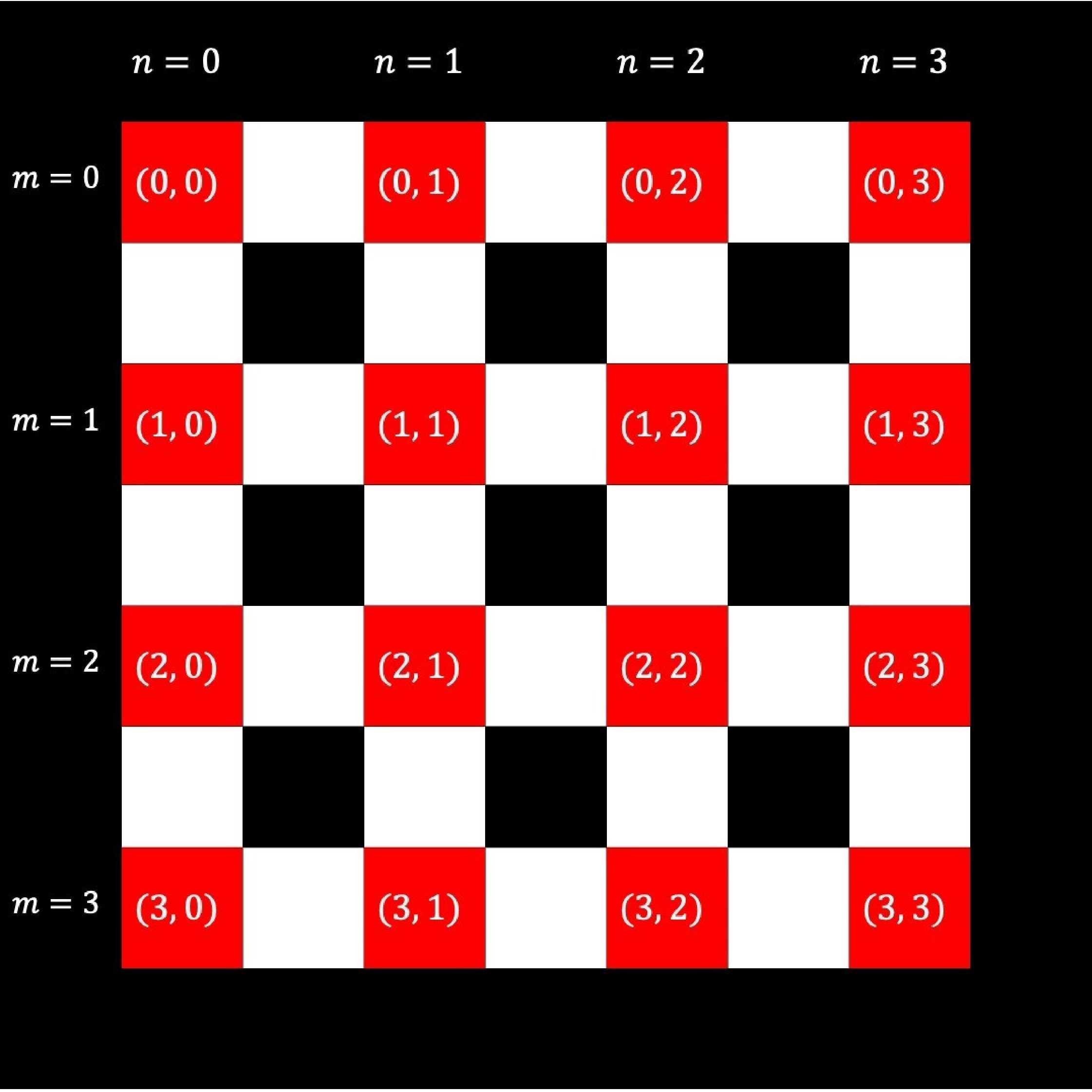}
\end{center}
\caption{ Black cells represent outer walls and inner bars $(i,j)$. Red cells represent options of start and goal coordinates $(m,n)$.
}\label{Bar-tipping_algorithm_start_goal_options}
\end{figure}
Figure \ref{Bar-tipping_algorithm_start_goal_options} shows the coordinates $(m,n)$ that are the options of the start and the goal. As Figure \ref{Bar-tipping_algorithm_start_goal_options} shows, $(m,n)$ is different from the coordinate setting bars; it is located at the four corners of the bars, where the bars do not extend. $X_{m,n}$ and $x_{i,j,d}$ are different. $X_{m,n}$ are options of start and goal, and $x_{i,j,d}$ are options of coordinates and directions to extend the bars.
We have shown the simplest implementation of the maze generation following the bar-tipping algorithm by quantum annealer.  Following the above, a maze, depending on randomness, is generated. To Generate a unique maze independent of randomness, we add the effect to make the maze more difficult in the cost function, and the difficulty is defined in terms of time (in seconds).

\subsection{Update rule}
We propose an additional $Q_{update}$ term to increase the time to solve the maze.
We introduce a random term that takes random elements to change the maze structure. It is added to the Equation \ref{cost_function}.
First, $Q_{update}$ term, the additional term which includes the new QUBO matrix $Q_{update}$, is given by
\begin{equation}
\begin{split}
\lambda_{update1}\sum_{i,i'}\sum_{j,j'}\sum_{d,d'}Q_{update(k,k')}x_{i,j,d}x_{i',j',d'} \\
+\lambda_{update1}\sum_{i}\sum_{j}\sum_{d}\sum_{m}\sum_{n}Q_{update(k,l)}x_{i,j,d}X_{m,n}\\
+\lambda_{update1}\sum_{i}\sum_{j}\sum_{d}\sum_{m}\sum_{n}Q_{update(l,k)}X_{m,n}x_{i,j,d}\\
+\lambda_{update2}\sum_{m,m'}\sum_{n,n'}Q_{update(l,l')}X_{m,n}X_{m',n'}, \\
\end{split}
\label{update}
\end{equation}
where
\begin{equation}
\left\{
\begin{array}{ll}
k=d+ (3N+1)i&(j=0 ) \\
k=d+3j+1 + (3N+1)i&(j\neq0)\\
l = (3N+1)N+(N+1)m + n.\\ 
\label{Q_update_subscripts}
\end{array}
\right.
\end{equation}
\begin{figure}[H]
\begin{center}
\includegraphics[width=10cm]{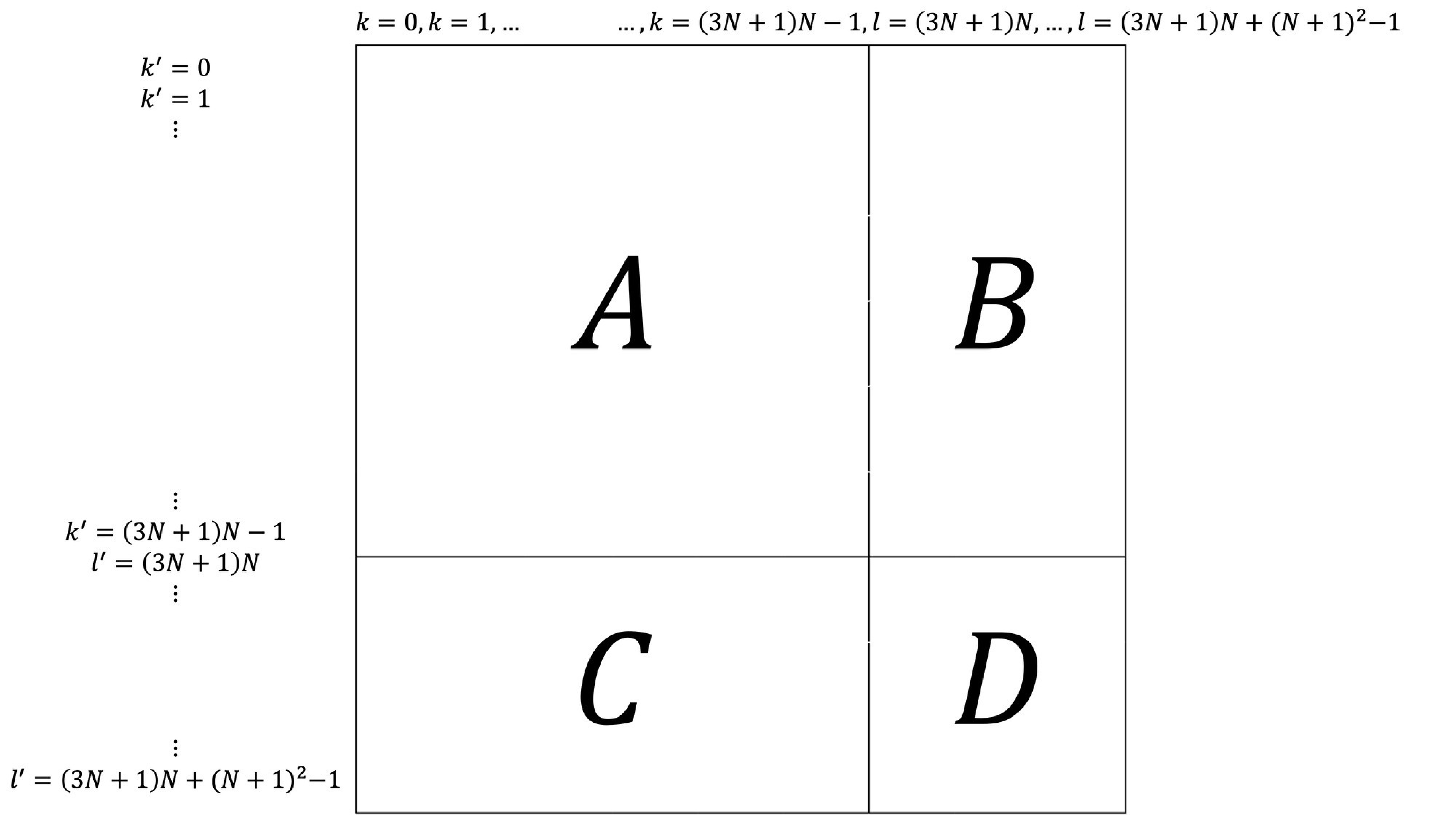}
\end{center}
\caption{Structure of $Q_{update}$. Part A is related to maze generation. Part B and part C are related to the relation between maze generation and the start and goal determination. Part D is related to the start and goal determination.  }\label{Qupdate_structure}\label{Maze}
\end{figure}
Figure \ref{Qupdate_structure} shows the structure of $Q_{update}$ and roles. 
Here, $k',l'$ are the replacement of $i,j,m,n$ in $k,l$ with $i',j',m',n'$. 
$N$ in Equation \ref{Q_update_subscripts} is the size of the maze.
The coefficients $\lambda_{update1}$ and $\lambda_{update2}$ are constants to adjust the effect of terms.
The elements of $Q_{update}$ related to maze generation, part A in Figure \ref{Qupdate_structure} is multiplied by the $\lambda_{update1}$. The elements of $Q_{update}$ related to the relation between the start and goal determination and the maze generation, part B, C in Figure \ref{Qupdate_structure} is multiplied by the $\lambda_{update1}$. The elements of $Q_{update}$ related to the start and goal determination, part D in Figure \ref{Qupdate_structure} is multiplied by the $\lambda_{update2}$.
These are to control the maze difficulty without breaking the bar-tipping algorithm's constraints. 
Equation \ref{update} is represented by the serial number $k$ of each coordinate $(i, j)$ at which bars can extend, and the sum $l$ of the total number of coordinates at which the bars can extend and the serial number of coordinates $(m, n)$, which are options for the start and the goal.
Furthermore, The second term and the third term in Equation \ref{update} allows the maze to consider the relation between the structure of the maze and the coordinates of the start and the goal.\\
    Second, $Q_{update}$, the new QUBO matrix, is given by
\begin{equation}
Q_{update} := p(t)Q_{update} + \bigl\{1-p(t)\bigr\}Q_{random},
\label{Q_update}
\end{equation}
where $Q_{random}$ is a matrix of random elements from $-1$ to $1$ and $p(t)$ depends on time $t$ (in seconds) taken to solve the previous maze and is expressed as follows
\begin{equation}
p(t)=\frac{1}{1+e^{-at}}.
\label{p(t)}
\end{equation}

The $Q_{update}$ is a matrix that was made with the aim of increasing the maze solving time through the maze solving iteration.
The initial $Q_{update}$ used in the first maze generation is a random matrix, and the next $Q_{update}$ that is used in the second or subsequent maze generation is updated using Equation\ref{Q_update}, the maze solving time $t$, and the previous $Q_{update}$.
The longer the solving time $t$ of the maze is, the higher the percentage of the previous $Q_{update}$ in the current $Q_{update}$ and the lower the percentage of $Q_{random}$; inversely, when $t$ is small, the ratio of the previous $Q_{update}$ is small, and the percentage of $Q_{random}$ is significant. 
In other words, the longer the solving time $t$ of the previous maze, the more characteristics of the previous term $Q_{update}$ remain. 
Here, $a$ is a constant to adjust the percentage. 
The $p(t)$ is a function that increases monotonically with $t$ and takes $0$ to $1$.
Thus, $Q_{random}$ that is, the random elements in $Q_{update}$ increase as time $t$ increases.
After the maze is solved, the next maze QUBO is updated by Equation \ref{Q_update} using the time taken to solve the maze. The update is carried out only once before the maze generation. Repetition of the update will make the maze gradually difficult for individuals.

The sum of Equation \ref{cost_function} and Equation \ref{update} is always used to generate a new maze annealing from a maximally mixed state.


\subsection{Experiments}
\subsubsection{Generation of maze}
    We generate mazes by optimizing the cost function using DW\_2000Q\_6. 
    Since the generated maze will not be solved, the update term is excluded for this experiment. \(\lambda_1 = 2\) and \(\lambda_2 = 2\) were chosen.

\subsubsection{Computational cost}
    We compare the generation times of \(N \times N\) maze in DW\_2000Q\_6 from D-Wave, simulated annealer called SASampler and simulated quantum annealer called SQASampler from OpenJij, D-Wave's quantum-classical hybrid solver called hybrid\_binary\_quadratic\_model\_version2 (hereinafter referred to as "Hybrid Solver") and classical computer (MacBook Pro(14-inch, 2021), OS: macOS Monterey Version 12.5, Chip: Apple M1 Pro, Memory: 16GB) based on bar-tipping algorithm coded with Python 3.11.5 (hereinafter referred to as "Classic"). 
    The update term was excluded from this experiment.
    We set \(\lambda_1 = 2\) and \(\lambda_2 = 2\).
    DW\_2000Q\_6 was annealed 1000 times for 20{\textmu}s, and its QPU annealed time for maze generation as calculated using time-to-solution (TTS).
    SASampler and SQASampler were annealed with 1000 sweeps.
    These parameters were constant throughout this experiment.
    Regression curves fitted using least squares method were drawn from the results to examine the dependence of computation time on maze size.

\subsubsection{Effect of update term}
    The solving time of \(9 \times 9\) maze generated without \(Q_{update}\) and using \(Q_{update}\) were measured. 
    This experiment asked 12 human subjects to solve mazes one set (30 times).
    To prevent the players from memorizing maze structure, they can only see the limited \(5 \times 5\) cells. 
    In other words, only two surrounding cells can be seen. 
    The increase rate from the first step of simple moving average of ten solving times was plotted on the graph.
    For this experiment, \(\lambda_1 = 2\), \(\lambda_2 = 2\), \(\lambda_{update1} = 0.15\), \(\lambda_{update2} = 0.30\) and \(a = 0.05\) were chosen. For two $\lambda_{update}$, we chose larger values that do not violate the constraints of the bar-tipping algorithm. We chose a value in which Equation \ref{p(t)} will be about 0.8 (80\%) when $t=30$ seconds as a constant $a$.

\subsection{Applicatons}
    The cost function in this paper has many potential applications by generalizing it.
For example, it can be applied to graph coloring and traffic light optimization.
Graph coloring can be applied by allowing adjacent nodes to have different colors.
Traffic light optimization can address the traffic light optimization problem by looking at the maze generation as traffic flow.
Roughly speaking, our cost function can be applied to the problem of determining the next state by looking at adjacent states.

    $Q_{update}$ can be applied to the problem of determining the difficulty of the next state from the previous result.
The selection of personalized educational materials is one of the examples. 
Based on the solving time of the previously solved problems, the educational materials can be selected at a difficulty suitable for the individual.
This is the most fascinating direction in future studies.
 As described above, we should emphasize that $Q_{update}$ proposed in this paper also has potential use in various fields related to training and education.

\section{Results}
 \subsection{Generation of maze}
    Figure \ref{Maze} shows execution examples of \(9 \times 9\) and \(15 \times 15\) mazes generated by optimizing the cost function using DW\_2000Q\_6. 

\begin{figure}[h!]
\begin{center}
\includegraphics[width=10cm]{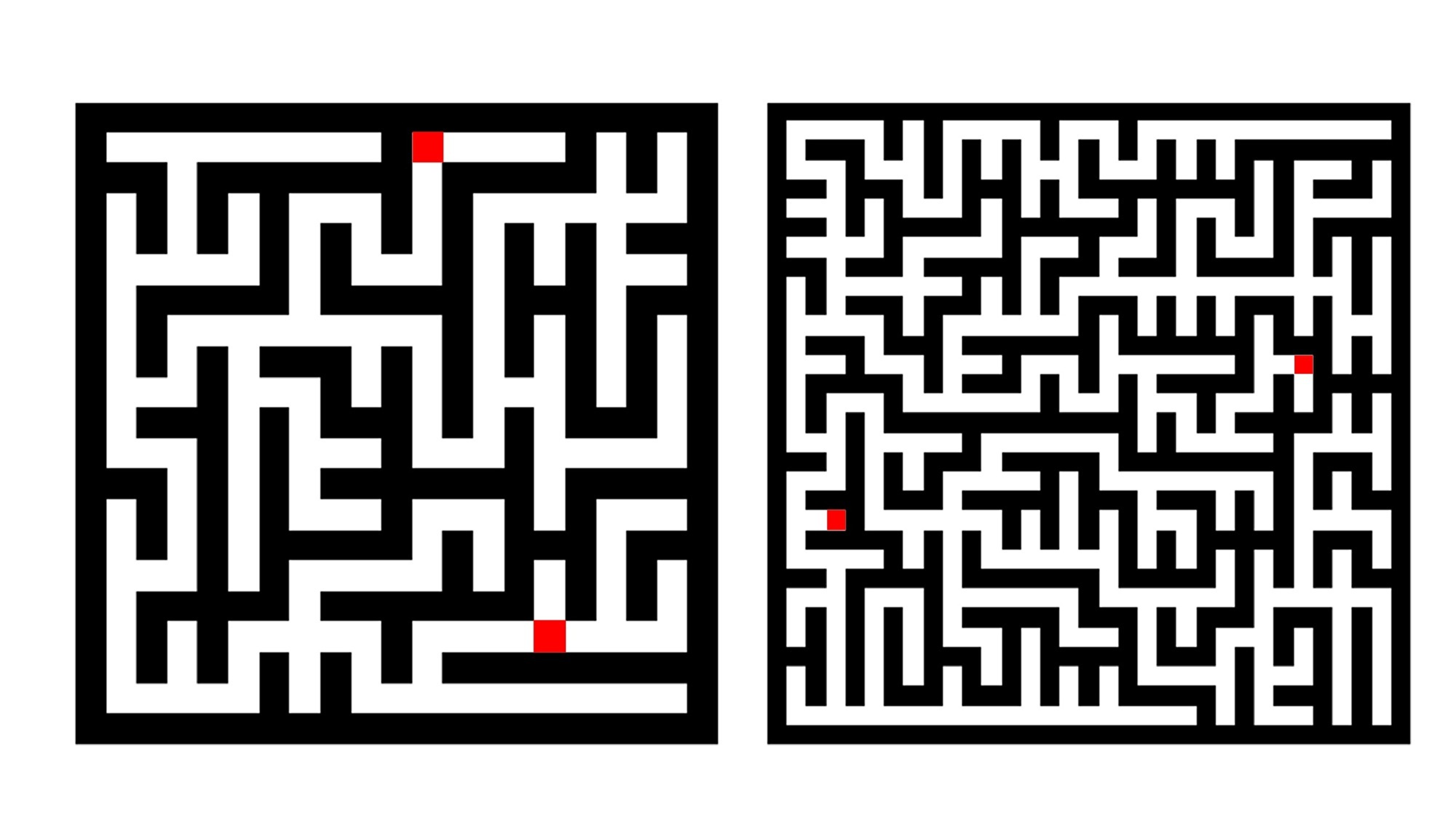}
\end{center}
\caption{ Left: \(9 \times 9\) maze generated by DW\_2000Q\_6. Right: \(15 \times 15\) maze generated by DW\_2000Q\_6.
Red cells represent a start and a goal for the maze.}\label{Maze}
\end{figure}

\subsection{Computational cost}

\begin{figure}[H]
\begin{center}
\includegraphics[width=10cm]{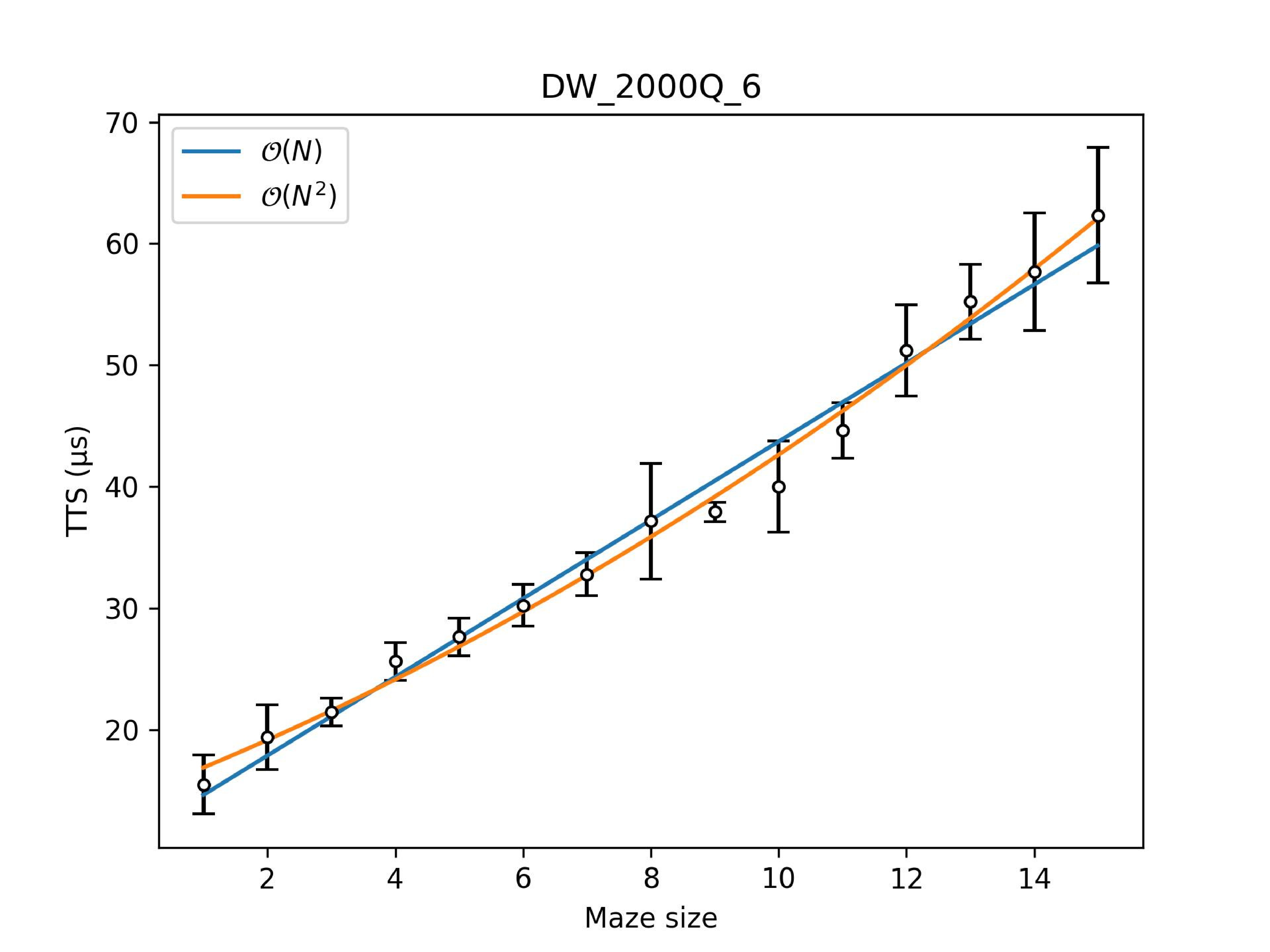}
\end{center}
\caption{ Time to reach the ground state with 99\% success probability as a function of the maze size in DW\_2000Q\_6. The error bars represent a 95\% confidence interval. The regression curve is given by \(\bigl((3.231 \pm 0.076)N + (11.40 \pm 0.69)\bigr)\) for linear regression and \(\bigl((7.4 \pm 1.8) \cdot 10^{-2} N^2 + (2.05 \pm 0.30)N + (14.8 \pm 1.0)\bigr)\) for quadratic regression.}\label{DW_2000Q_6_fitting}
\end{figure}

\begin{figure}[H]
\begin{center}
\includegraphics[width=18cm]{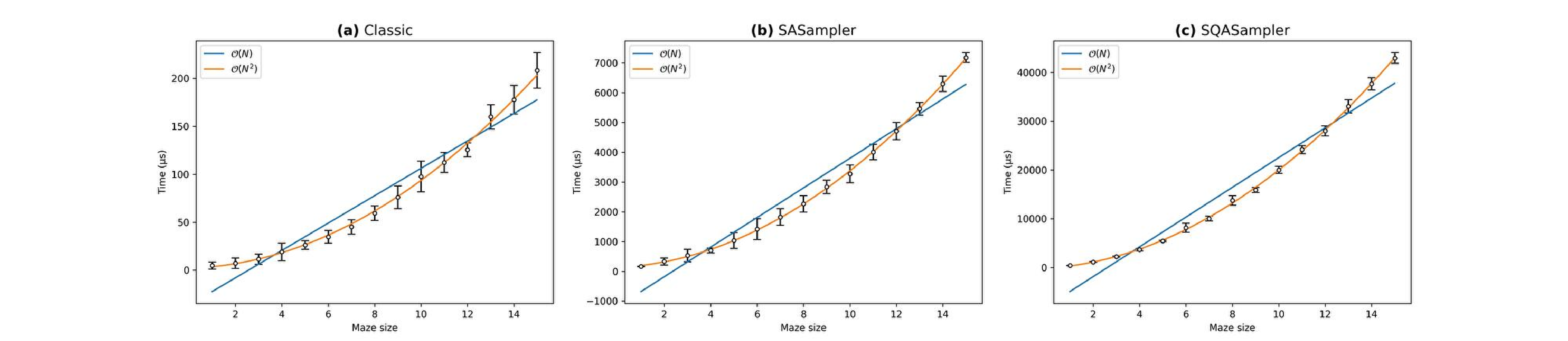}
\end{center}
\caption{ \textbf{(a)} The time to reach the ground state as a function of the maze size in Classic. The error bars represent a 95\% confidence interval. The regression curve is \(\bigl((0.855 \pm 0.090)N^2 + (0.6 \pm 1.5)N + (2.2 \pm 5.1) \bigr)\). \textbf{(b)} Time to reach the ground state as a function of the maze size in SASampler. The error bars represent a 95\% confidence interval. The regression curve is \(\bigl((28.8 \pm 1.2)N^2 + (36 \pm 20)N + (129 \pm 71)\bigr)\). \textbf{(c)} Time to reach the ground state as a function of the maze size in SQASampler. The error bars represent a 95\% confidence interval. The regression curve is \(\bigl((172.8 \pm 4.4)N^2 + (287 \pm 73)N -(1.5 \pm 2.5) \cdot 10^{2} \bigr)\)}\label{Classic_solvers_fitting}
\end{figure}


\begin{figure}[H]
\begin{center}
\includegraphics[width=10cm]{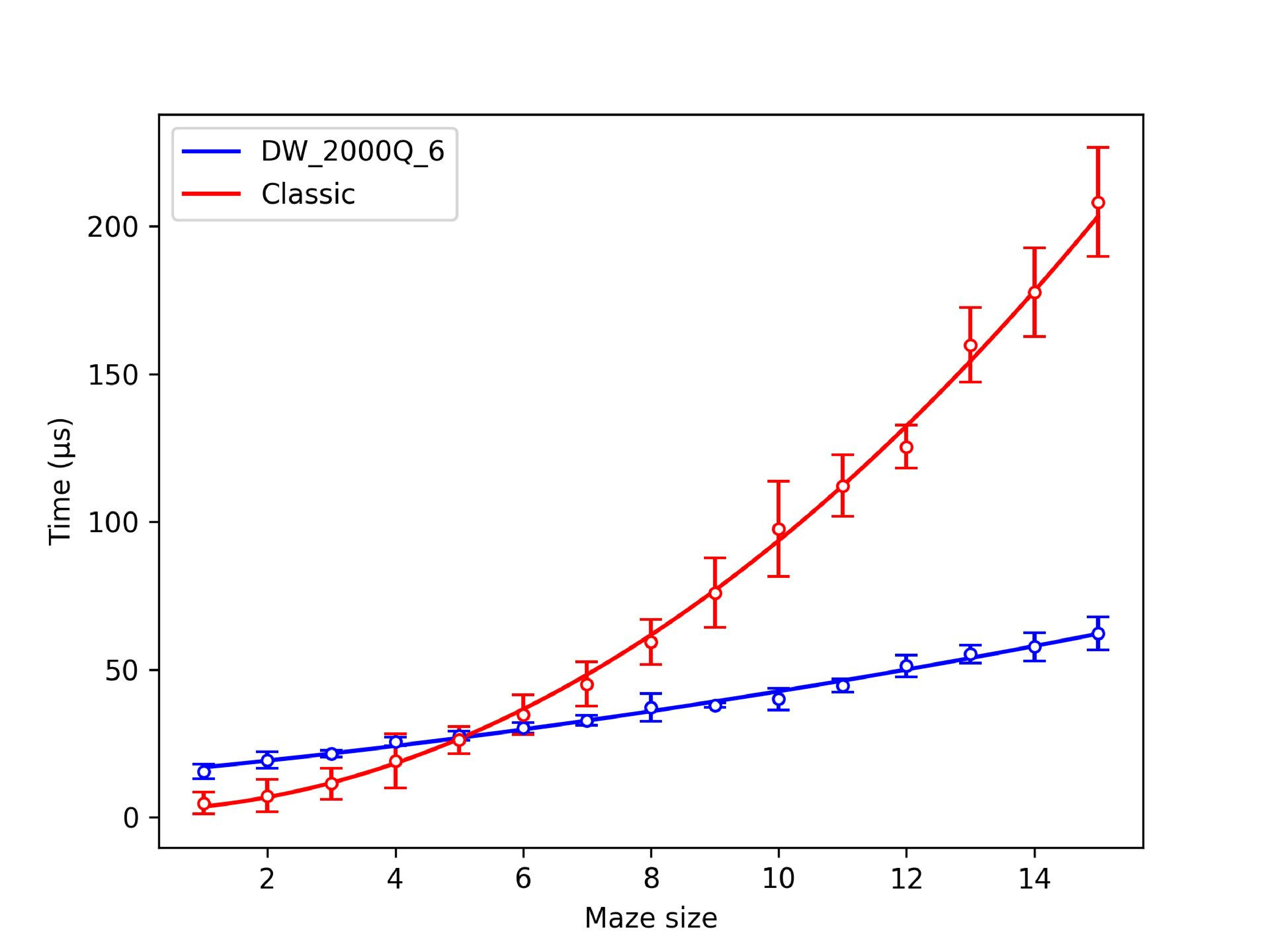}
\end{center}
\caption{ Comparison of maze generation time between DW\_2000Q\_6 and Classic.}\label{DW_2000Q_6_Classic_fitting_comparison}
\end{figure}

\begin{figure}[H]
\begin{center}
\includegraphics[width=10cm]{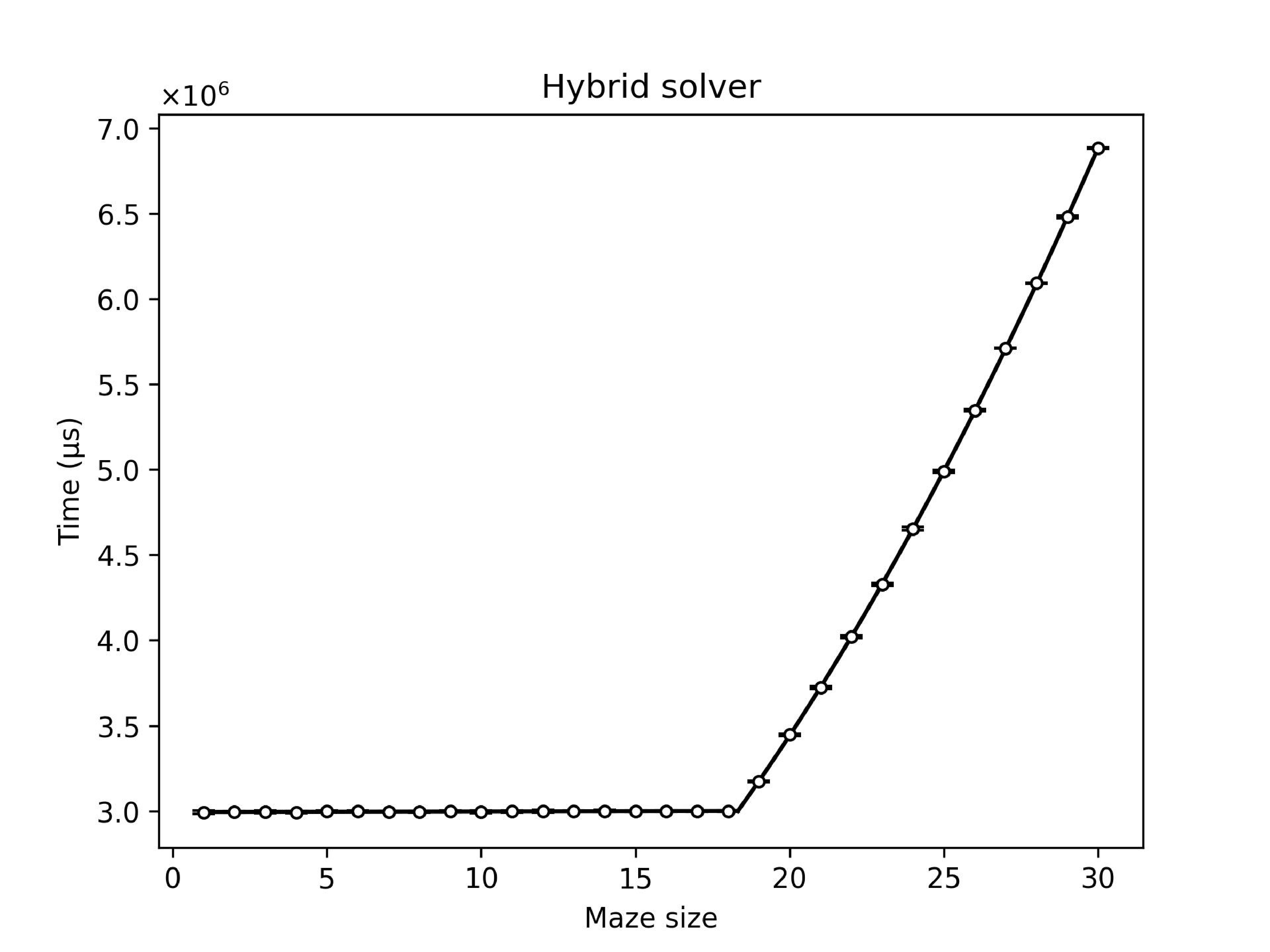}
\end{center}
\caption{ Time to reach the ground state as a function of maze size in the Hybrid Solver. The error bars represent a 95\% confidence interval.}\label{Hybrid_solver_fitting}
\end{figure}
    Fits of the form $aN^2 + bN + c$ are applied to each of the datasets using least squares method. The results are as follows.
    Figure \ref{DW_2000Q_6_fitting} shows the relation between TTS for maze generation and maze size on DW\_2000Q\_6. DW\_2000Q\_6 is \(\mathcal{O} (N)\) or \(\mathcal{O} (N^2)\). 
    Even if it is quadratically dependent on the maze size, its deviation is smaller than the other solvers. 
    Figure \ref{Classic_solvers_fitting} shows the relation between maze generation time and maze size on Classic, SASampler, and SQASampler. Classic $\bigl((0.855 \pm 0.090)N^2 + (0.6 \pm 1.5)N + (2.2 \pm 5.1) \bigr)$, SASampler $\bigl((28.8 \pm 1.2)N^2 + (36 \pm 20)N + (129 \pm 71)\bigr)$, and SQASampler $\bigl((172.8 \pm 4.4)N^2 + (287 \pm 73)N -(1.5 \pm 2.5) \cdot 10^{2} \bigr)$ exhibit quadratic dependence on the maze size \(\mathcal{O} (N^2)\). 
    Most of the solvers introduced here are \(\mathcal{O} (N^2)\) since they are extending $N \times N$ bars to generate a maze.
    Figure \ref{DW_2000Q_6_Classic_fitting_comparison} shows the comparison of maze generation time between DW\_2000Q\_6 and Classic. DW\_2000Q\_6 has a smaller coefficient $N^2$ than the classical algorithm, and after $N=5$, DW\_2000Q\_6 shows an advantage over Classic in the maze generation problem. 
    The improvement using quantum annealing occurred because it determines the direction of $N \times N$ bars at once.
    Figure \ref{Hybrid_solver_fitting} shows the relation between maze generation time and maze size on Hybrid Solver. 
    Linear and quadratic fits applied to the dataset indicate the Hybrid Solver is \(\mathcal{O}(1)\) or \(\mathcal{O}(N)\) \(\bigl((3.29 \pm 0.83) \cdot 10^{2}N + (2.99325 \pm 0.00090) \cdot 10^{6}) \bigr)\) between \(N = 1\) and \(N = 18\) and then shifted to \(\mathcal{O}(N^2)\) \(\bigl((6.899 \pm 0.065) \cdot 10^{3}N^2 -(0.4 \pm 3.2) \cdot 10^{3}N + (6.90 \pm 0.39) \cdot 10^{5}\bigr)\).
    The shift in the computational cost of Hybrid Solver may have resulted from a change in its algorithm.

\subsection{Effect of update term}

\begin{figure}[h!]
\begin{center}
\includegraphics[width=15cm]{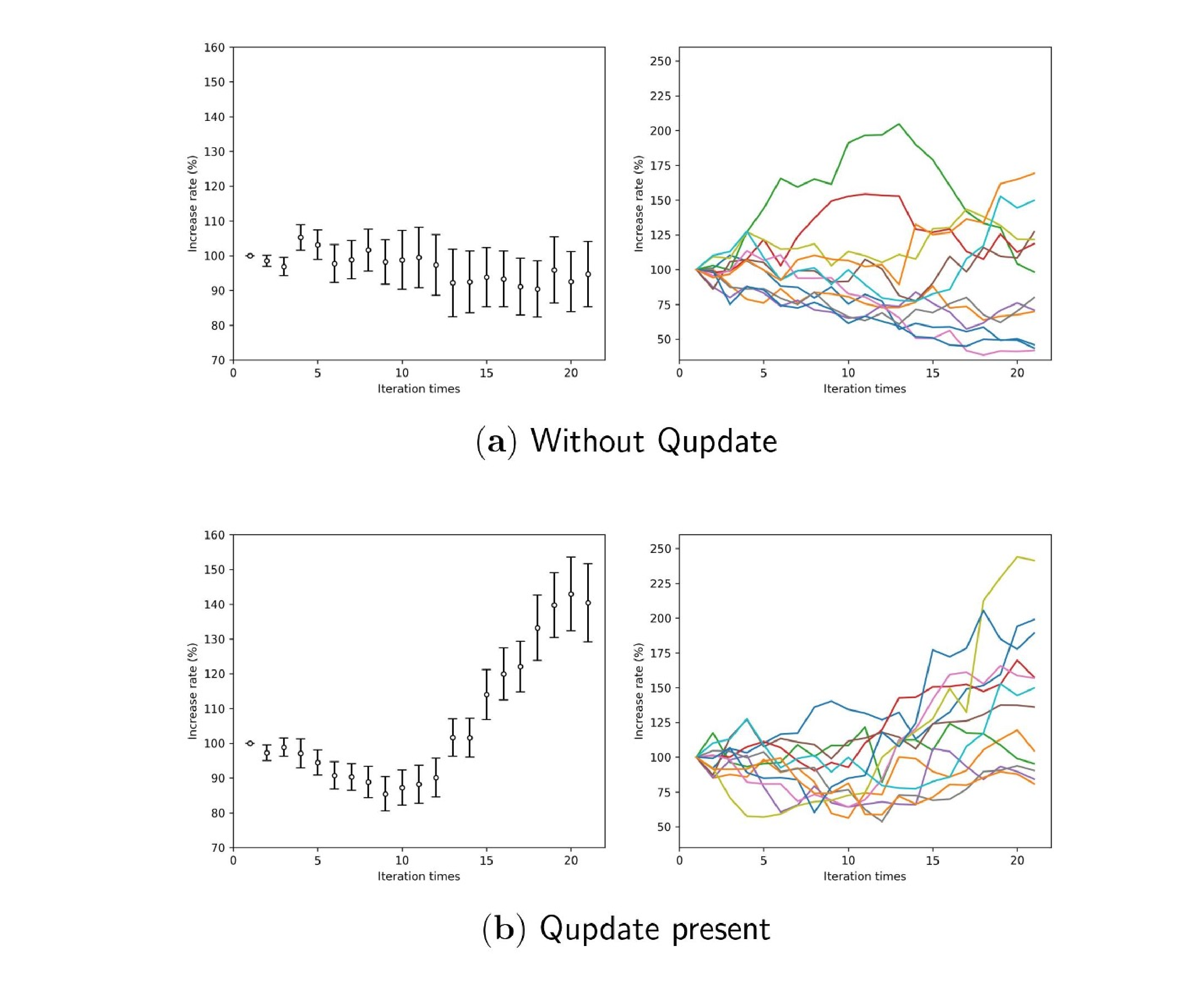}
\end{center}
\caption{ \textbf{(a)} Left: Increase rate from the first step of simple moving average of 10 solving time of \(9 \times 9\) maze generated without \(Q_{update}\). The error bars
represent standard errors. Right: All players' increase rate from the first step of simple moving average of 10 solving time of \(9 \times 9\) maze generated without \(Q_{update}\). \textbf{(b)} Left: Increase rate from the first step of simple moving average of 10 solving time of \(9 \times 9\) maze generated using \(Q_{update}\). The error bars represent standard errors. Right: All players' increase rate from the first step of simple moving average of 10 solving time of \(9 \times 9\) maze generated using \(Q_{update}\). }\label{Solvingtime}
\end{figure}   

    Here, 12 human subjects are asked to solve the maze one set (30 times), and the maze is shown to increase in difficulty as it adapts to each human subject.
    Figure \ref{Solvingtime} (a) shows the increase rate from the first step of simple moving average of 10 solving time of maze generated without $Q_{update}$ and individual increase rate. 
    The solving time of the maze without \(Q_{update}\) was slightly getting shorter overall. 
    Figure \ref{Solvingtime} (b) shows the increase rate from the first step of simple moving average of 10 solving time of maze generated using $Q_{update}$ and individual increase rate.
    The solving time of the maze using \(Q_{update}\) was getting longer overall. 
    Most of the players increased their solving time, but some players decreased or didn't change their solving time.
    In addition, nine players' average of the solving time of the maze generated using \(Q_{update}\) increased than that of the maze generated without \(Q_{update}\).
    These show that $Q_{update}$ has potential to increase the difficulty of the mazes. 

\section{Discussion}
In this paper, we show that generating difficult (longer the maze solving time) mazes using the  bar-tipping algorithm is also possible with quantum annealing.
By reformulating the bar-tipping algorithm as the combinatorial optimization problem, we generalize it more flexibly to generate mazes.
In particular, our approach is simple but can adjust the difficulty in solving mazes by quantum annealing.

In Sec.3.2, regarding comparing computational costs to solve our approach to generating mazes using TTS, DW\_2000Q\_6 has a smaller coefficient of $N^2$ than the classical counterpart. 
Therefore, as $N$ increases, the computational cost of DW\_2000Q\_6 can be expected to be lower than that of the classical simulated annealing for a certain time.
Unfortunately, since the number of qubits in the D-Wave quantum annealer is finite, the potential power of generating mazes by quantum annealing is limited.
However, our insight demonstrates some advantages of quantum annealing against its classical counterpart.
In addition, we observed that the hybrid solver's computational cost was constant up to $N=18$.
This indicates that hybrid solvers will be potentially effective if they are developed to deal with many variables in the future. 

In Sec. 3.3, we proposed $Q_{update}$ to increase the solving time using quantum annealing. 
We demonstrated that introducing $Q_{update}$ increased the time to solve the maze and changed the difficulty compared to the case where  $Q_{update}$ was not introduced. 
At this time, the parameters ($\lambda_{update1}$, $\lambda_{update2}$, and $a$) were fixed.
Difficult maze generation for everyone may be possible by adjusting the parameters individually.

One of the directions in the future study is in applications of our cost function in various realms.
We should emphasize that $Q_{update}$ proposed in this paper also has potential use in various fields related to training and education.
The powerful computation of quantum annealing and its variants opens the way to such realms with high-speed computation and various solutions.

\section*{Conflict of Interest Statement}
Sigma-i employs author Masayuki Ohzeki. 
The remaining authors declare that the research was conducted without any commercial or financial relationships that could be construed as a potential conflict of interest.

\section*{Author Contributions}
Y. I. , T. Y. , and K. O. conceived the idea of the study. M. O. developed the statistical analysis plan, guided how to use the quantum annealing to find the optimal solution, and contributed to interpreting the results. 
Y. I. , T. Y. , and K. O. drafted the original manuscript. 
M. O. supervised the conduct of this study. 
All authors reviewed the manuscript draft and revised it critically on intellectual content. 
All authors approved the final version of the manuscript to be published.

\section*{Funding}
The authors thank financial support from the MEXT-Quantum Leap Flagship Program Grant No. JPMXS0120352009, as well as Public\verb|\|Private R\&D Investment Strategic Expansion PrograM (PRISM) and programs for Bridging the gap between R\&D and the IDeal society (society 5.0) and Generating Economic and social value (BRIDGE) from Cabinet Office.

\section*{Acknowledgments}
The authors thank the fruitful discussion with Reo Shikanai and Yoshihiko Nishikawa on applications of our approach to another application.
This paper is the result of research developed from an exercise class held at Tohoku University in Japan in the past called "Quantum Annealing for You, 2nd party!".
We want to thank one of the supporters, Rumiko Honda, for supporting the operations.
The participants were a diverse group, ranging from high school students to university students, graduate students, technical college students, and working adults. 
As you can see from the authors' affiliations, this is a good example of a leap from the diversity of the participants to the creation of academic and advanced content.

\bibliographystyle{Frontiers-Harvard} 
\bibliography{bibdata}

\end{document}